# Exposure and remediation in the French maritime sector: a passive measurement of 129 organisations

Edouard Vallet, Colomban Monnier

VAIATA Dynamics, Rians, France · edouard.vallet@vaiata-dynamics.fr, colomban@vaiata-dynamics.fr

**Abstract**

Maritime transport and ports are essential entities under the EU NIS2 directive, but there is little public data on the external exposure of the sector in France. In March 2026, after physical attacks on Western interests in the Middle East raised the risk of cyber retaliation against French maritime operators, we assessed the internet-facing attack surface of 129 organisations: 103 French companies and 26 international operators active in French waters. The March assessment was passive. It used DNS and email-authentication records, TLS certificates, search-engine indexing, archived pages, public JavaScript, cloud storage naming and third-party internet-scan data (Shodan), with no connection attempt beyond ordinary public web requests. After manual verification we retained 475 findings, 3.7 per organisation. 46% of organisations had at least one high or critical finding and 11% at least one critical finding, including exposed databases, unpatched services with known CVEs, an exposed credentials file, publicly listed cloud buckets and a subdomain open to takeover. The most common gap was email spoofing protection, with no enforced DMARC policy at 60% of organisations. French companies fared worse than international operators (49% vs 35% at high or critical risk; 4.0 vs 2.3 findings per organisation). Between 2 and 18 March 2026, individual reports were sent for 43 of the audited domains. We rescanned 123 organisations in November 2026. Among March critical and high findings that could be rechecked, 52% were gone for organisations that had received a report and 54% for those that had not, so private email disclosure had no measurable effect on remediation. Results are aggregated and anonymised.

## 1. Introduction

The maritime sector's exposure to cyber attacks is documented by incidents at large carriers and by regulation: the International Maritime Organization requires cyber risks to be addressed in safety management systems (Resolution MSC.428(98)) [1], and NIS2 (Directive (EU) 2022/2555) classifies maritime transport operators and ports as essential entities with risk-management and incident-notification duties [2].

Public measurements of the exposure of a national maritime sector are rare. Industry surveys rely on self-reporting, and vendor reports seldom publish their method or their population. We ask two questions. How exposed is the sector, seen from the internet? And does a private report make an organisation fix what was found? We answer the first with a reproducible, passive measurement of 129 organisations operating in France in March 2026, broken down by finding type, severity and origin (French vs international operators). We answer the second by measuring the same organisations again in November 2026 and comparing those that received a private report with those that did not. We ran the measurement in March because physical attacks on Western interests in the Middle East had made retaliation by state-aligned actors against French and European maritime interests a plausible scenario, and passive reconnaissance shows what such an actor would see first.

## 2. Related work

Maritime cyber risk has mostly been studied through threat models and risk frameworks rather than measurement; MaCRA [4] is a representative model-based framework for ships and maritime systems. Large-scale vulnerability notification has been measured outside the maritime sector. After Heartbleed, notified operators patched faster than a control group [5]. Li et al. [6] found that notifications sent to direct contacts improved remediation, while repeated notices did not. Stock et al. [7, 8] showed that email notification of website owners reaches few recipients and produces modest effects. Section 6 tests the same question on a small, sector-specific population where reports were sent by name to identified contacts.

## 3. Population

Organisations were drawn from the member lists of French maritime professional bodies (Armateurs de France, GASPE coastal ship-owners association, Cluster Maritime Français, Union des Ports de France), supplemented with major international operators with activity or interests in France. We kept 129 organisations with analysable web infrastructure: ship-owners, ports, ferry and cruise operators, towage and mooring cooperatives, port services, offshore, classification societies, maritime software vendors and energy shippers. 103 are headquartered in France and 26 abroad.

## 4. Method

For each organisation we identified primary domains and subdomains from public sources and ran thirteen passive modules:

| Module family | Sources and checks |
|---|---|
| DNS and email | SPF, DMARC and DKIM records; subdomain enumeration from public sources such as certificate transparency |
| TLS | certificate validity, expiry, hostname match |
| HTTP | security headers, technology fingerprint from ordinary page loads |
| Internet-scan data | Shodan records: open ports, service banners, CVEs attached to detected versions |
| Search-engine indexing | targeted queries ("dorks") for indexed documents, error pages, admin interfaces |
| Archives | Wayback Machine snapshots of sensitive paths |
| JavaScript | public scripts scanned for embedded keys and endpoints |
| Cloud and takeover | public bucket naming and listing; dangling DNS records to unclaimed cloud resources |
| Breach data | public breach indexes for organisational domains |

Table 1. Passive modules. Port and CVE data come from Shodan's existing records.

Every candidate finding was verified manually. We removed false positives such as CDN and WAF artefacts, soft-404 pages, intentionally public keys and deliberately published scientific data directories. Each finding was rated low, medium, high or critical according to the access it would give an attacker (for example, a reachable database port or credentials in a public file is critical). Each organisation is rated by its most severe finding.

## 5. Results (March 2026)

### 5.1 Overall exposure

We retained 475 findings: 20 critical, 68 high, 207 medium and 180 low, a mean of 3.7 per organisation. Table 2 gives the distribution of organisations by their most severe finding.

| Highest severity | Organisations | Share |
|---|---:|---:|
| Critical | 14 | 11% |
| High | 45 | 35% |
| Medium | 51 | 40% |
| Low | 7 | 5% |
| None detected | 12 | 9% |

Table 2. Organisations by most severe finding (n = 129).

### 5.2 Finding categories

| Category | Findings | Share |
|---|---|---|
| Email authentication (SPF, DMARC, DKIM) | 178 | 37% |
| Indexed leaks (search engines) | 130 | 27% |
| Exposed ports and known CVEs | 114 | 24% |
| TLS | 22 | 5% |
| Exposed files | 18 | 4% |
| Manual findings | 6 | 1% |
| Secrets in JavaScript | 3 | 1% |
| Cloud storage | 3 | 1% |
| Subdomain takeover | 1 | 0% |

Table 3. Findings by category (n = 475).

The most frequent individual findings, as a share of organisations, were: DKIM absent 39%, DMARC policy p=none 31%, DMARC absent 29%, permissive SPF (~all) 27%, port 8080 exposed 17%, port 8443 exposed 13%, FTP exposed 12%, invalid certificate 12%, SPF absent 12%, known CVEs on exposed services 10%, SMTP exposed 8%, certificate about to expire 5%. Counting absent records and p=none, 60% of organisations had no enforced DMARC policy, so their domains can be spoofed.

### 5.3 Critical findings

The 14 organisations rated critical include a cruise company, a maritime software vendor, a major port, ship-owners, a gas carrier, a mooring cooperative and its professional body, a classification society, an oil major and a river-and-sea transport group. The critical findings were: three organisations with a database port (MySQL 3306 or PostgreSQL 5432) reachable from the internet; thirteen servers with publicly documented CVEs attached to their detected service versions; a ship-owner serving a configuration (.env) file with API keys in clear text; cloud buckets with public listing; a subdomain pointing to an unclaimed cloud resource, open to takeover; and video-surveillance equipment of a major port reachable from the internet.

### 5.4 French companies and international operators

| | French (103) | International (26) |
|---|---|---|
| Findings | 416 | 59 |
| Findings per organisation | 4.0 | 2.3 |
| High or critical risk | 49% | 35% |
| Critical organisations | 10 | 4 |

Table 4. French companies vs international operators.

French organisations, many of them small and medium coastal operators, show almost twice as many findings per organisation as the large international groups. The likely explanation, not tested here, is size and dedicated security teams.

### 5.5 Regulatory reading

89% of organisations had at least one finding that is a potential indicator of NIS2 non-compliance (unmanaged vulnerabilities, unprotected communications, exposed critical services).

## 6. Remediation after eight months

On 15 and 16 November 2026 we reran the same thirteen modules on the same domains, with the same false-positive rules. Six organisations were not rescanned because our pre-flight checks excluded them, leaving 123. A March critical or high finding counts as fixed when the module that produced it ran in both scans and no longer reports it; findings on modules that did not run for a domain in March are not compared.

| | Report sent | No report |
|---|---:|---:|
| Domains rescanned | 39 of 43 | 84 of 86 |
| March critical/high findings rechecked | 44 | 35 |
| Fixed | 23 (52%) | 19 (54%) |
| Fixed, excluding certificates close to expiry | 21 of 42 (50%) | 14 of 30 (47%) |
| New critical/high findings | 8 | 10 |

Table 5. Remediation of March critical and high findings, by report status.

About half of the serious findings disappeared in eight months whether or not the organisation had received a report. Among organisations that had, 21 findings were still present in November, including FTP and MySQL ports open to the internet, services with known CVEs and three cloud buckets that could still be listed. The last "Fixed" row excludes certificates close to expiry, which automatic renewal fixes without any action.

As in [7, 8], a report sent once by email, without follow-up or a sector authority behind it, had no measurable effect on remediation. The difference between groups is within noise, and the groups are small (44 and 35 findings), so a modest effect cannot be excluded.

Two differences between the runs affect the comparison. Shodan records reflect each month's crawl, so a port missing in November can be an expired record rather than a closed port, and search-engine queries went through a different provider in November. A single November observation cannot date a fix, so some remediation may come from routine maintenance or later contacts. The list of reports comes from copies of the sent mail and is a lower bound, so the no-report group may include recipients.

During the November run the scanner also opened direct TCP connections, which we had not intended, to nine database and remote-access ports on about 150 addresses. The 84 open ports it found are excluded from all figures, the probe was removed once found, and the March data contain no such result.

## 7. Discussion

Most findings are ordinary, cheap-to-fix weaknesses in email authentication and service exposure. Shipping operations rely on email between ship-owners, agents, brokers, ports and authorities, where a spoofed instruction (berth change, payment details, crew change) has direct operational impact. Exposed services with known CVEs and credentials in public files give initial access without any phishing.

The retaliation against French maritime interests that motivated the study did not materialise as a visible cyber campaign during the study period. Had it done so, a substantial share of the sector offered low-effort entry points.

## 8. Limitations

Third-party scan records can be stale, and known-CVE findings rely on version banners. Passive data confirm neither exploitability nor NIS2 non-compliance. The severity rating is ours, not a standard score. The sample is purposive, limited to organisations with analysable web infrastructure, and over-represents organised, visible actors, so it is not a statistical estimate for all French maritime companies. Ship systems and operational technology are out of scope.

## 9. Ethics and disclosure

No intrusion, authentication attempt or exploitation was performed, in line with French law (Code pénal, art. 323-1 to 323-7); the unintended November connection attempts are disclosed in Section 6. Between 2 and 18 March 2026, before any publication, individual reports with findings and remediation advice were sent for 43 of the audited domains, covering 40 of the 129 organisations; in three cases the report went to the parent group or to a sector body rather than to the audited entity. No organisation or domain is named in this paper. A summary of the study was published by Le marin on 4 April 2026 [3].

## 10. Conclusion

A passive assessment of 129 maritime organisations operating in France produced 475 verified findings, with 46% of organisations at high or critical risk. Email spoofing protection was missing at 60% of organisations, and a small set exposed databases, credentials or vulnerable services. Eight months later, about half of the March critical and high findings had been fixed, at the same rate whether or not the organisation had received a report. A single private e-mail did not change remediation in this sector. The method needs no access or cooperation, and can be repeated to measure the effect of other disclosure channels.